# Unique shape changes during freezing of a nanofluid droplet

Yugang Zhao[‡], Chun Yang[*,‡]

[‡]School of Mechanical and Aerospace Engineering, Nanyang Technological University, 50 Nanyang Avenue, Singapore 639798

*E-mail: mcyang@ntu.edu.sg; Tel: (+65) 6790-4883

## Abstract:

Understanding the dynamics during freezing of nanofluid droplets is of importance from a fundamental and practical viewpoint. While the freezing of a water droplet has been extensively studied, little information is available about the characteristics of freezing a nanofluid droplet. Here, we report unique shape changes observed during the freezing process of a nanofluid droplet. Instead of forming a pointy tip on the frozen deionized water droplet[1], we found that the top of a frozen nanofluid droplet exhibits a flat plateau shape and such plateau becomes larger with increasing nano-particle concentration. We ascribe this characteristic shape change to an outward Marangoni flow which moves liquid from the droplet's interior to its vertexes, thus engendering the volume redistribution and shape changes. Moreover, we also observed a particle ring pattern formed during freezing of a nanofluid droplet, which is similar to the well-known "coffee stain" phenomenon occurring during evaporation of a water droplet with dispersed particles[2]. We believe our reported phenomena are not only of fundamental interest but have potential applications in freezing of nanofluids for energy storage and developing nanofluid based new microfabrication methods.

Liquid solidification in droplet form has been a classic model system widely used for understanding of heat transfer with phase change phenomena. In particular, the freezing of a water droplet on a cold plate has been studied intensively over decades[3, 4, 5, 6, 7, 8, 9], due to its frequent occurrence in nature and many application processes. Existing studies have shown the complex physics and unorthodox behaviors of water during solidification[10, 11, 12, 13, 14]. Intriguingly, a unique behaviour is that the frozen water droplet expands , yielding a tip singularity despite the presence of liquid surface tension[1]. Moreover, several interesting phenomena were reported; the frost halos formed from freezing an impact subcooled water droplet[15], the explosion of a partially frozen droplet[16], and the two overlapping stages during freezing of a sessile droplet[17] are a few to name.

Study the freezing of nanofluid (i.e., liquid with dispersed nanoparticles) droplets is crucial to a broad range of technological contexts such as nanostructure-enhanced phase change materials for thermal energy storage[18, 19], cryobiology and food engineering[20, 21, 22, 23], and soil remediation[24, 25] etc. Specifically, recent studies on the directional freezing of colloidal suspensions have led to developing the so-called "ice-templating" method[26, 27], which can be used in a versatile processing route for the fabrication of bioinspired porous materials and composites[28, 29]. To date, how the presence of nanoparticles will affect single droplet freezing processes has not been reported in the literature yet.

Here we report a new discovery that the final shape of a frozen nanofluid droplet changes with the concentration of nanoparticles. Instead of forming a pointy tip as in the case of a pure water droplet[1], the top of a nanofluid ice droplet exhibits a flat plateau shape and such plateau is found to become larger with increasing particle concentration. To explain unique shape changes during freezing of a nanofluid droplet, we develop an analytical model to illustrate the physics associated with this phenomenon. We ascribe this characteristic shape change to an outward Marangoni flow which moves liquid from droplet interior to its vertexes during the propagation of freezing front, resulting in the droplet volume redistribution and shape change. Additionally, we observe a particle ring pattern formed near the plateau edge during the freezing of a nanofluid droplet; such phenomenon is similar to the well-known "coffee stain" phenomenon occurring during evaporation of a droplet with particle dispersions[2, 30].

We carried out a direct visualization experiment to study the freezing process of single nanofluid droplet on a smooth silicon wafer substrate that was maintained at a constant subcooled temperature 20 ºC. More details about the preparation of nanofluids and substrate surface and the experimental setup will be described in the Methods section.

Figure 1a shows a schematic of the experimental setup. A deionized (DI) water or nanofluid droplet of its volume about 6.7 ± 0.4 µL (with a typical radius of $R_0 \approx 1$ mm) was first generated from a quartz capillary and then was gently released to deposit on the subcooled substrate (due to gravitational force). Nanofluid droplets, consisting of 15 nm titanium oxide particles of various particle concentrations dispersed in DI water containing 0.8 mM dissolved cationic surfactant, Hexadecyltrimethylammonium bromide, have slight volume variation with particle concentration (Supplementary Figure 1a).

The droplets started freezing upon in contact with the substrate surface as a result of the heterogeneous nucleation occurring spontaneously at the liquid/solid substrate interface. Figures 1b-1e show the extracted images of completely frozen droplets having various particle volumetric concentrations. For a deionized water droplet without any particles (Figure 1b), the resultant ice droplet is a truncated sphere with a pointy tip which has been already reported in the literature[1, 31], resulted from the expansion of water upon freezing[1]. The intricate part, however, is a flat plateau formed on the top of ice droplets for the case of nanofluids (Figures 1c-1e). The plateau size increases dramatically with increasing the nanoparticle volumetric concentration from 0.1% to 0.5%. The equivalent "contact angle" of the ice droplet was $\theta = 63.5 \pm 5^o$, and it was not dependent on particle concentration. Also, despite slightly higher thermal diffusivity and viscosity of the nanofluid[32, 33, 34, 35], all droplets got completely frozen within no more than 3 seconds, showing that the droplet freezing time is not noticeably dependent on particle concentration (Supplementary Figure 1b). We found the formation of such flat plateau for frozen nanofluid droplets is universal, because as shown in Supplementary Figure 2, the phenomenon also occurred under other experimental conditions, including the silicon wafer substrates with different contact angle (wettability) and the nanofluid droplets with different volume, nanoparticle size and particle material.

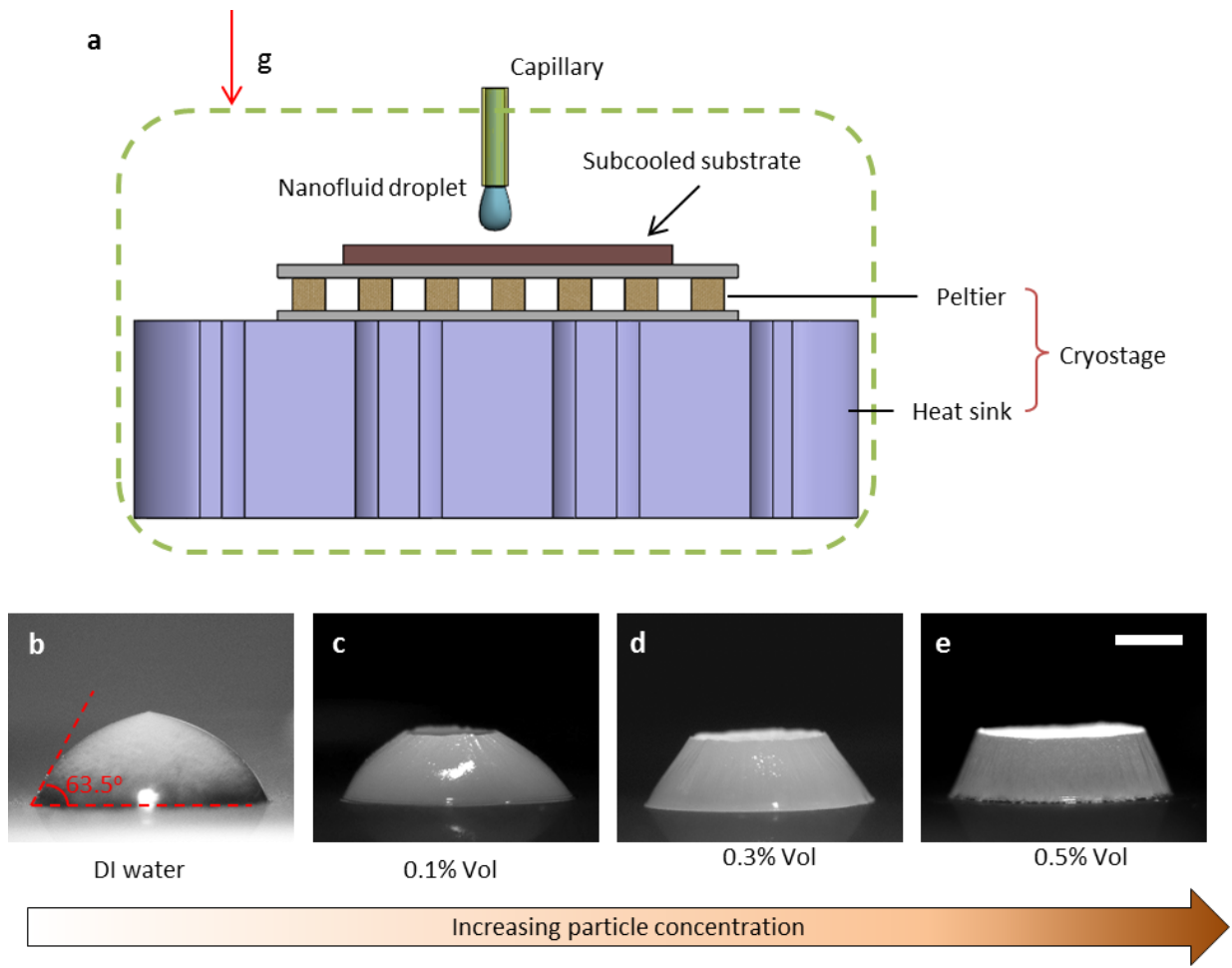

**Figure 1│ In-situ observation of the shape of ice droplets. a**, Schematic depiction of the cryostage used for observing the freezing process of a droplet deposited on a smooth silicon wafer substrate with a fixed subcooled temperature of -20 ºC. **b-e**, Optical images showing the final shapes of frozen deionized water droplet and 15 nm Titanium Oxide nanofluid droplets with three different particle volumetric concentrations of 0.1%, 0.3% and 0.5%. The scale bar denotes 1 mm in length.

We also made a comparison of the freezing dynamics between a DI water droplet (Supplementary Movie 1) and a nanofluid droplet that has a moderate particle volumetric concentration of 0.3% (Supplementary Movie 2). Figures 2a and 2b respectively show the selected snapshots during the freezing processes of a DI water droplet and a 0.3 % v/v nanofluid droplet from four different stages, including: at the beginning of freezing, the freezing front approaching half of the frozen droplet height, right before and after completion of the freezing. We digitalized these images and the digitalized profiles of the DI water and 0.3 % v/v nanofluid droplets are presented in Figures 2c and 2d, respectively. For the DI water droplet, the droplet profile propagates upwards as the result of volume expansion of freezing water. In contrast, for the nanofluid droplet, the droplet profile moves downwards and even gets flattened after the freezing completes. Specifically, it is noted that there is a volume redistribution from the droplet apex to its vertexes.

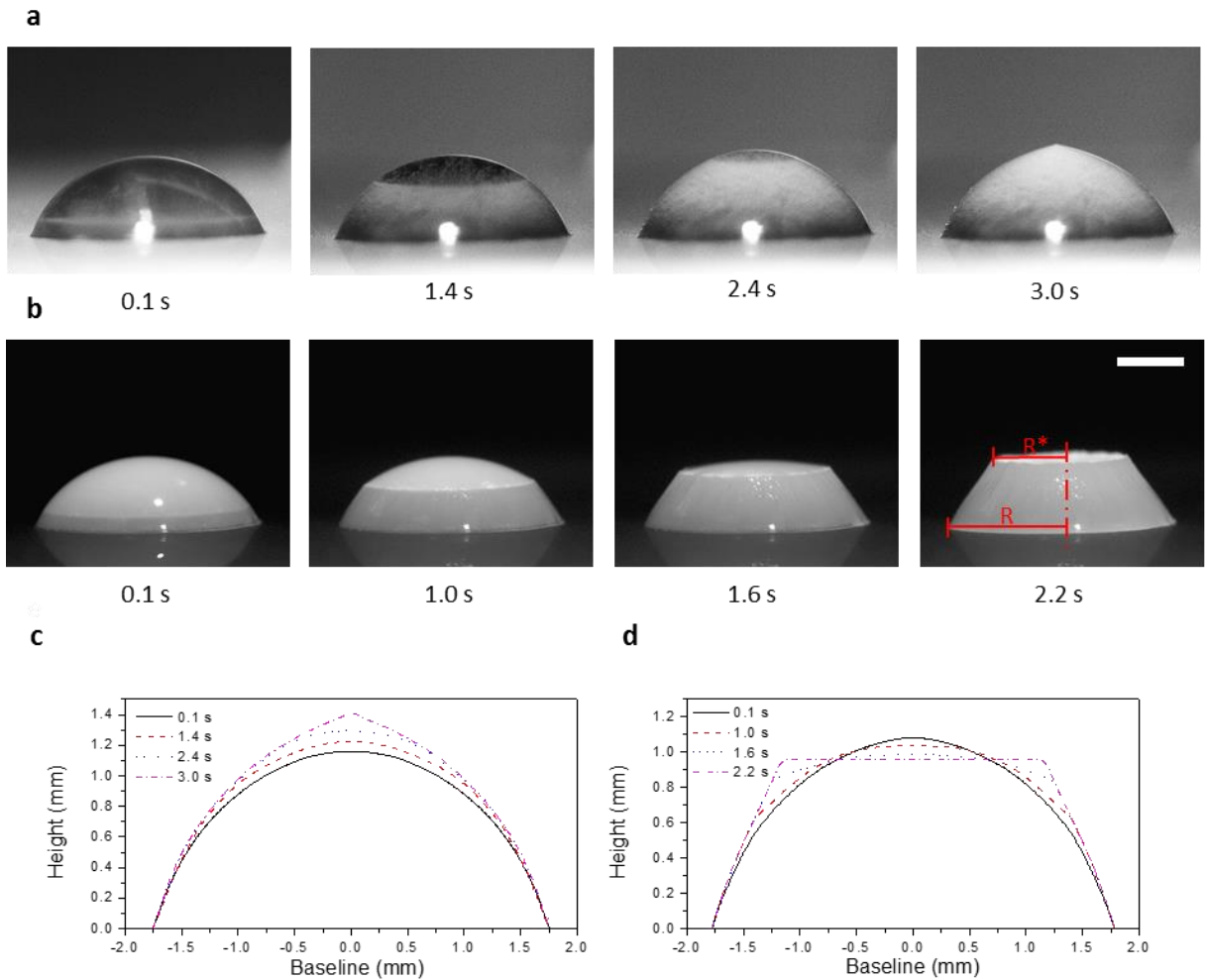

**Figure 2 | Freezing dynamics of a deionized water droplet and a nanofluid droplet. a**, Selected snapshots showing the freezing process of a deionized water droplet. **b**, Selected

snapshots showing the freezing process of a nanofluid droplet with a moderate particle volumetric concentration of 0.3 %. $R$ and $R^*$ are the droplet radius and half of the top plateau, respectively. **c**, **d**, The digitalized droplet profiles corresponding to **a** and **b**, respectively. Inserts show the droplet profiles inside the selected regions indicated by two dashed rectangles. The scale bar denotes 1 mm in length.

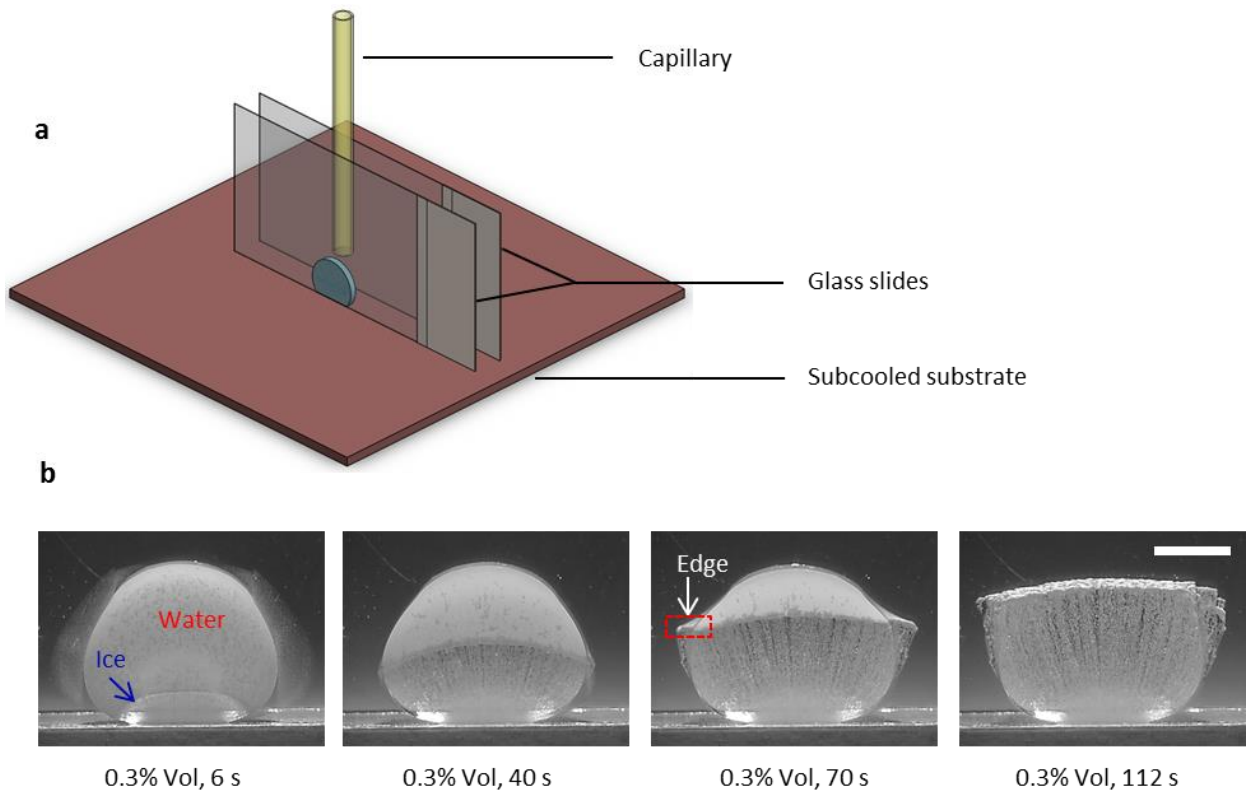

**Figure 3│Freezing dynamics of a nanofluid droplet in a Hele-Shaw cell. a**, Schematic depiction showing a Hele-Shaw cell mounted on a subcooled substrate. **b**, Selected snapshots of a freezing nanofluid droplet with a moderate particle volumetric concentration of 0.3% in the Hele-Shaw cell. The scale bar denotes 1 mm in length.

To eliminate 3D droplet curvature effects, we further investigated the freezing characteristics of a nanofluid droplet in a 2D Hele-Shaw cell. This allows to clearly observe the evolution of freezing front and the growth of top plateau. As illustrated in Figure 3a, the Hele-Shaw cell is mounted on the subcooled silicon wafer substrate and it was constructed using two microscope glass slides separated by a 1mm spacer and with a quartz capillary inserted between for generating and introducing a single nanofluid droplet. To prevent the optical distortion from formation of wetting meniscus on glass slides, the inner surfaces of the two glass slides were specifically pretreated with chemical vapor deposition of silane, resulting in a static contact angle of about 90º.

To demonstrate the reliability of our Hele-Shaw device, the dynamic a freezing DI water droplet in the Hele-Shaw cell is provided in Supplementary Figure 3 which shows clearly that we were able to reproduce the results reported by a recent literature study on the freezing of a DI water droplet[1]. Figure 3b shows the selected snapshots of a freezing nanofluid droplet with a moderate particle volumetric concentration of 0.3% in the Hele-Shaw cell. The ice and water phases can be readily detected from their distinguished optical properties, and thus the profile of the freezing front can be obtained (at 6s and 40 s). Since the heat transfer area from the subcooled substrate in the Hele-Shaw cell is much smaller than that in a 3D unconfined droplet case (Figure 2b), the freezing in the Hele-Shaw cell takes much longer time. Also a considerable amount of particles are found to be segregated as the freezing front propagates upwards and a particle enrichment zone is observed. Consequently, two sharp edges (on both left and right sides) emerge at a certain stage of freezing (at 70 s) and they grow into a large plateau on the top when the freezing completes (at 112 s). A more detailed dynamic process of the one edge growth is given in Supplementary Movie 3. The movie also provides a direct experimental evidence that there is a flow transporting the liquid from the nanofluid droplet interior to its vertexes. Figure 4a shows a snapshot of the edge growth in the preselected zone as highlighted in Figure 3b. Nanoparticles are observed to accumulate in the advancing part of the freezing front and particularly in the vicinity of vertexes where a higher particle concentration is expected. Figure 4b depicts a schematic illustration of Figure 4a to describe the resultant particle redistribution and a Marangoni flow. The reason of developing such Marangoni flow is as below. During the nanofluid droplet freezing process, a certain amount of nanoparticles is segregated in front of the freezing front, and the segregated particles are accumulated near the edge (Figure 3b and Figure 4a), giving rise to particle concentration variation along the droplet-air free interface. As the surface tension of nanofluids depends on the particle concentration[36, 37, 38] (see Supplementary Figure 6), a surface tension gradient is built up along the droplet-air free interface and hence a Marangoni flow is generated. This suggests that a considerable amount of nanoparticles segregated during the freezing front propagation is a critical precondition accounting for Marangoni flow development. Existing literature studies show that the segregation behavior is resulted from the interplay between the advancing freezing front and particles. Specially, the segregation coefficient is related to the velocity of advancing freezing front (freezing rate) and particle radius[39, 40]. Therefore, the manifestation of the shape change during the freezing of a nanofluid droplet depends on freezing rate and particle size.

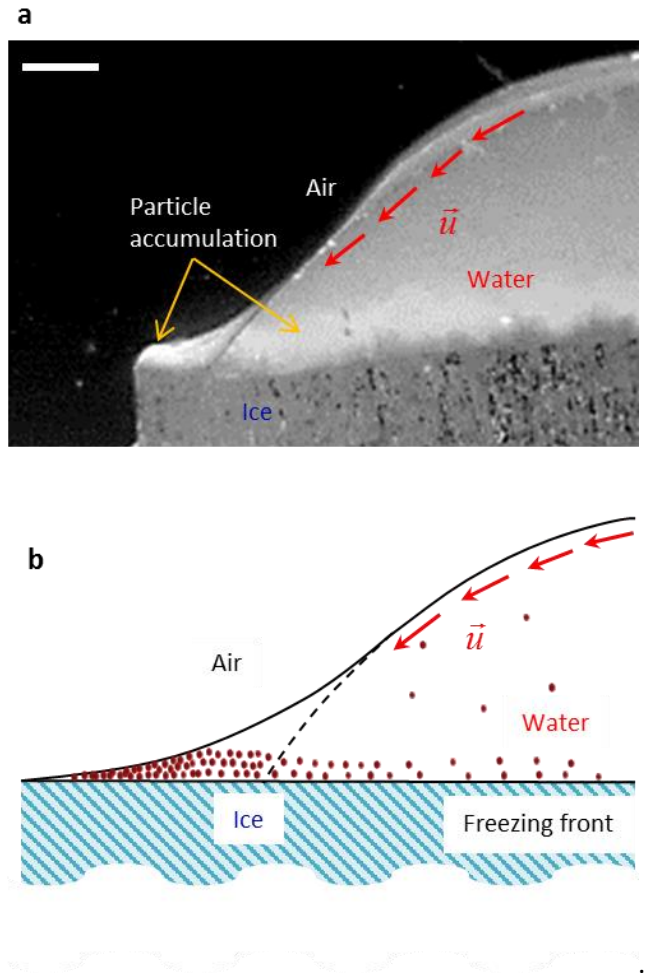

**Figure 4│An outward Marangoni flow induced by the particle concentration difference along the droplet-air free interface. a**, Direct observation of an outward Marangoni flow and the edge formation (indicated by a selected box in Figure 3b) during the freezing of a 0.3% v/v nanofluid droplet in a Hele-Shaw cell (see Supplementary Movie 3 for detailed dynamic process). The scale bar denotes 50 µm in length. **b**, Schematic illustration of the outward Marangoni flow (its direction indicated by solid arrows) as a result of particle concentration variation built up along the droplet-air free interface (with higher concentration near the edge due to the segregated particles accumulated there).

Additionally, we observed a particle ring pattern formed during freezing of a nanofluid droplet,

which is similar to the well-known "coffee stain" phenomenon occurring during evaporation of a water droplet with dispersed particles. Particles are accumulated near the vertexes as this outward Marangoni flow continuously bring the nanofluid from droplet interior while only solid particles are left and the liquid medium gets frozen. It is thus expected that these edge structures in the Hele-Shaw cell will convert into a ring pattern for a freezing nanofluid droplet on the smooth surface. A similar phenomenon occurs in the evaporation of a nanofluid droplet, referred to the classic "coffee ring effect"[2, 30, 41, 42, 43, 44]. where nanoparticles accumulate to the pinned boundary and form a ring pattern. Figure 5 shows a ring patterns formed in the freezing of a nanofluid droplet. The dynamic process of the ring formation in the freezing of a nanofluid droplet can be found in Supplementary Movie 4. In both of the two cases nanoparticles are accumulated to the vertexes and the aggregated particle assemblies form a ring pattern.

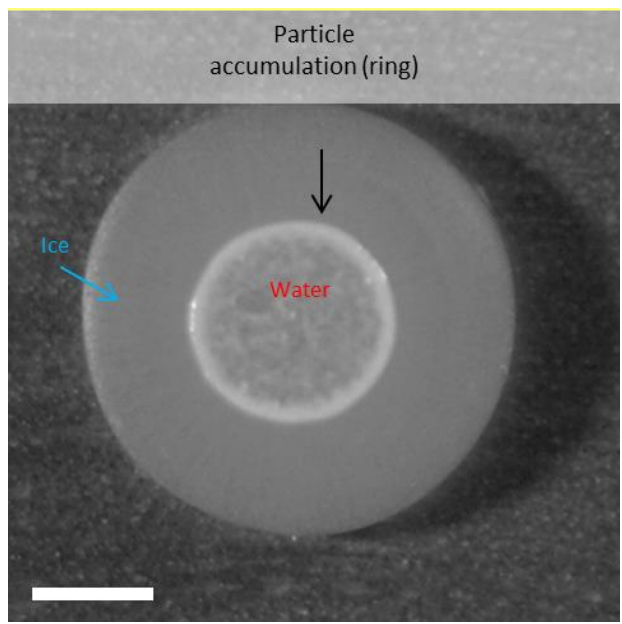

**Figure 5│Ring patterns formed from freezing and evaporating of nanofluid droplets.** A snapshot showing the formation of a ring pattern from freezing a nanofluid droplet in present work (white color referring to a higher concentration). The scale bar denotes 500 µm in length.

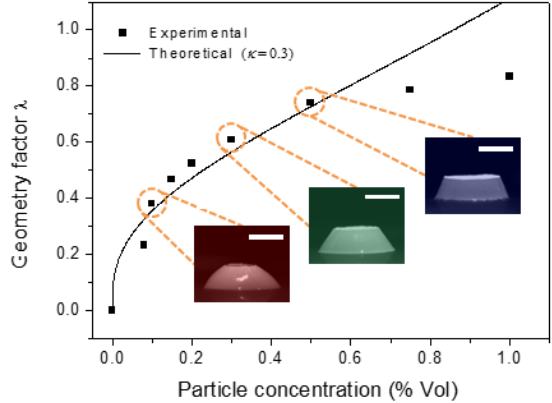

**Figure 6 | Dependence of the shape change (geometric factor $\lambda$) on the particle volume concentration.** Solid line shows theoretical predictions from our model and square scatters show the experimental measurements. Inserts show the corresponding experimental observations with false colors. The scale bar denotes 1 mm in length.

To quantitatively elucidate the effect of nanoparticles on the shapes of ice droplets, we developed a scaling model on the basis of the following three major assumptions: (1) an appreciable amount of nanoparticles are segregated during propagation of the freezing front, thus the particle concentration is different along the free interface; (2) the particle concentration difference causes the variation of surface tension, i.e., Marangoni stress, which induces an outward flow; (3) the flow transports liquid from the droplet interior to the vicinity of vertexes, thus engenders the shape change. These assumptions are proposed basing on our experimental observations, in particular as shown in Figure 4 and Supplementary Video 3. We describe the model in a 2D cylindrical coordinate as shown in Supplementary Figure 7. The droplet is treated as a spherical cap when neglecting the gravitational force as $R<Ca$, where $Ca=\sqrt{\gamma/\rho g}$ ~ 2.7 mm is the capillary length, with $\gamma$, $\rho$ and $g$ being the surface tension, the liquid density and the gravitational acceleration, respectively.

Firstly, we estimate the magnitude of the surface tension gradient along the free interface. The particle concentration $\phi(r,z)$ within the liquid residual is determined by the balance of freezing segregation and particle diffusion. An appreciable particle concentration difference occurs when $t_{fre} \ll t_{dif}$, where $t_{fre} \sim$ 2-3 s is the time needed to freeze a droplet, and $t_{dif} = R^2/D \sim 2.7 \times 10^4$ s is the time scale for mass diffusion. $D$ is the diffusivity of particles computed through the Stokes-Einstein equation $D = \frac{k_B T}{6\pi\mu a}$, with $k_B$ denoting the Boltzmann constant, $T$ the absolute temperature, $\mu$ the dynamic vicosity of the base fluid, and $a$ the radius of nanoparticles. We express the particle volume concentration difference from the droplet vertex to the droplet top along the free interface as $\Delta\phi \sim (C-1)\phi_0$, where $C \equiv \phi_f/\phi_0$ is the segregation coefficient, with $\phi_0$ denoting the initial particle volume concentration, and $\phi_f$ is the particle volume concentration at the advancing freeing front. For a colloidal system with monodispersed hard spherical particles, we express its surface tension as[45] $\gamma = \gamma_0 + f(a)\phi$, where $\gamma_0$ is the surface tension of the base fluid (water), and $f(a)$ is a geometric factor. Applying a linear approximation, we express the magnitude of the surface tension gradient along the free interface as, $\nabla_S \gamma \sim \frac{2f(C-1)\phi_0}{\pi r_0}$, where $\vec{s} = \left(\vec{e}_r + \frac{\partial z}{\partial r}\vec{e}_z\right)/\sqrt{\left(\frac{\partial z}{\partial r}\right)^2 + 1}$ is the tangential unit vector, and $r_0$ is the base radius of the liquid residual. Secondly, we compute the resultant flow through a force balance between the Marangoni stress and the viscos stress expressed as, $\mu\left(\frac{\partial u_r}{\partial z} + \frac{\partial u_z}{\partial r}\right) = \nabla_S \gamma$, where $u_r$ and $u_z$ are the velocity components along horizontal and vertical directions. As the shape change is mainly attributed to a mass transfer in horizontal direction, we herein only consider the horizontal component of this flow. We eliminate $u_z$ from the force balance equation, leading to an expression as $\mu\left(1 - \frac{r^2}{z^2}\right)\frac{\partial u_r}{\partial z} + \mu\frac{u_r}{z} = \nabla_S \gamma$. Combining the relation of surface tension and particle concentration, we obtain a simple expression as $u_r \sim \dfrac{f(C-1)\phi_0}{r_0 \mu}$. Thirdly, we compute the volume of transported fluid by integrating this horizontal flow over the freezing time, $\Delta V = \int_{t_0}^{t_0+t_{fre}} J dt$, where $J = \bar{u}_r 2\pi r h$ is the volume transfer rate at an arbitrary location $(r, h)$ and $\bar{u}_r \equiv \dfrac{1}{h}\int_0^h u_r dz$ is the average velocity of the horizontal flow over the height of the liquid residual. Our recent work[46] has shown $t_{fre} \sim P(R,\theta)\dfrac{L}{k\Delta T}$, where $P(R,\theta)$ is a geometric factor, $L$ is the latent heat of fusion, $k$ is thermal conductivity of ice, and $\Delta T$ is the degree of subcooling. We can thus express a dimensionless form of the transported fluid as $\Delta V/V_0 = E(a,C,\phi_0) M(\mu,L,k,\Delta T) N(R,\theta)$, where $V_0$ is the volume of droplets, $E(a,C,\phi_0)$ represents the effect of nanoparticles, $M(\mu,L,k,\Delta T)$ represents the effect of the base fluid properties and substrate temperature, and $N(R,\theta)$ represents the effect of liquid droplet morphology. Specially, $E(a,S,\phi_0) = f(C-1)\phi_0$ shows that the dimensionless form of the transported fluid changes linearly with the particle volume concentration. Fourthly, we quantify the shape change during freezing of nanofluid droplets and calibrate its relationship with the transported liquid. We introduce a dimensionless geometric factor $\lambda \equiv R^*/R$, where $R^*$ is the droplet radiuses at the top plateau as shown in Figure 2b. A geometric analysis has been conducted as shown in Supplementary Figure 8, leading to the following expression, $\Delta V/V_0 = \left[1 - 1/(\lambda^2 + \lambda + 1)\right]^3$. Accordingly, setting $\kappa = \dfrac{MN}{f(C-1)}$ as the induced coefficient, representing effects of all other parameters except for the particle volume concentration $\phi_0$, we express the geometric factor $\lambda$ as a function of the particle volume concentration,

$$\left[1 - 1/\left(\lambda^2 + \lambda + 1\right)\right]^3 = \kappa \phi_0.$$

To examine the predictive capacity of our model, we measured the shapes of ice droplets from DI water ($\phi_0=0$) to a maximum volume concentration of 1% experimentally. Figure 6 shows the comparison between our experimental results and our model predictions where $\kappa \approx 0.3$ is the fitting value using the least squares method. Generally, our model can capture the trend of our experimental results very well, suggesting that this shape changes during freezing nanofluid droplets can be explained through our theoretical model. However, the theoretical predictions markedly differ from experimental results for the concentration of $\phi_0 \approx 0.8\%$ or above. The reason could be that segregated particles form a band with a constant packing density at relatively high concentrations[47, 48]. Therefore, the effective particle concentration difference along the free interface and the resultant Marangoni flow is thus kept relatively unchanged afterwards.

In summary, we reported a new discovery in the frame of phase change of nanofluids, i.e. unique shape changes from freezing nanofluid droplets. Instead of forming a pointy tip as the case of a deionized water droplet, a frozen nanofluid droplet exhibits a flat plateau on the top and such plateau becomes larger with increasing particle concentration. Correspondingly, the shape of an ice droplet can be altered from a tip singularity, to a truncated cone and a circular cylinder by judicious control of the nanoparticle concentration. We envisage that this versatile and robust strategy to control the frozen droplet shape in a highly repeatable manner promises advances in high-resolution 3D printing, and microfabrication technologies. This work may be readily extended to a variety of materials from bio-fluids, polymers to metallic materials for specific applications.

The present finding on its mechanism reveals that an outward Marangoni flow continuously moves liquid from droplet interior to vertexes and causes this shape change. A scaling model basing on that understanding agrees well with our experimental measurements. From a boarder perspective, the study on the freezing of nanofluid droplets represents an important advance in our understanding of the phase change in complex systems. Specially, from comparing the cases of freezing and evaporating nanofluid droplets, we show that the redistribution of nanoparticles from a homogeneous state to a heterogeneous state (the formation of a ring pattern) during the phase change seems to be universal for a colloidal system, inspiring more efforts to reveal their correlations.

## Methods

**Nanofluid preparation.** In our experiment, the nanofluid was prepared by mixing well-degassed deionized water (Millipore, 18 mΩ-cm) and titanium oxide nanoparticles (US Research Nanomaterials, anatase) at a variety of particle volume concentrations from 0.01% to 1%. Those nanoparticles are nearly monodispersed, consisting of particles with an approximately spherical morphology and their diameters were characterized using a field emission scanning electron microscope (FESEM, JEOL7600) as shown in Supplementary Figure 9. A cationic surfactant Hexadecyltrimethylammonium bromide (CTAB, Sigma-Aldrich) was added to prevent agglomeration of nanoparticles and promote system stability at a fixed concentration of 0.8 mM (critical micelle concentration of 1 mM at room temperature). After stirring mixing for 30 mins, we kept prepared nanofluids in the ultrasonic chamber (Elmasonic, 37 kHz) for more than 2 hours to break possible large particle aggregations and remove dissolved gas. The characterization of nanofluid stability can be found in Supplementary Figure 4, 5. With titanium oxide having an albedo near unity, the suspensions resulted in an opaque white liquid, similar to milk.

**Substrate characterization.** A smooth silicon based substrate was used in this work. Single side polished single crystal silicon wafers were cut into squares of size 2 cm × 2 cm with a diamond cutter. Chemical vapor deposition (CVD) of a monolayer of trichloro (1H, 1H, 2H, 2H-perfluorooctyl) silane (Sigma-Aldrich) was applied to promote hydrophobicity. Firstly, the substrate was rinsed with acetone, isopropyl alcohol and deionized water subsequently, and dried in nitrogen gas flow. Then, the substrate was functionalized with hydrophilic groups under oxygen plasma (Harrick plasma) for 10 mins, then immediately placed into a vacuum chamber containing an open container of silane at room temperature for 30 mins. Upon removal from the chamber, the substrate was rinsed again with IPA and DI water and then dried under nitrogen gas flow. We characterized the substrate wettability by depositing a 5 μL DI water droplet onto the surface and measuring its static contact angle. The results from contact angle measurement (Attension Theta) showed that a static contact of $\theta = 47^o$ for a bare silicon substrate and $\theta = 105^o$ for a silanized substrate, respectively (seen in Supplementary Figure 9). During the experiment, the substrate was attached onto the cryostage using thermal paste with polished side upwards.

**Experimental setup.** An isolated cryostage together with a precise thermal control unit, a droplet delivery system, and a high-speed visualization system, as shown schematically in Supplementary Figure 10, was used to study the freezing of nanofluid droplets in this work.

The cryostage consisted of a heat sink and a Peltier element (Ferrotec, 9500/127/060B) to achieve a dual-stage control of substrate temperature. The heat sink was made by connecting a hollow aluminum block to a cooling circulator (JULABO, bath fluid thermal-H5). A layer of Teflon covering the heat sink helped minimizing the heat loss and achieving lower subcooled temperature. The Peltier element which was powered by a programmable DC power supply was bonded onto the heat sink via thermal paste to minimize heat loss due to contact thermal resistance. On top of the Peltier element was a sample substrate bonded with the same thermal paste. Three T-type thermal couples were mounted between the sample substrate and the Peltier element to measure the substrate temperature. The substrate temperature was adjusted by varying the temperature and/or flowrate of the cooling circulator and the output of the Peltier element. During experiment, the substrate temperature was maintained at a constant temperature -20 ºC achieved by a proportional-integral-derivative controller via a LabVIEW data acquisition input & output module (National instrument, NI 9219 & NI 9264).

The isolated chamber was made of Teflon as the frame with acrylic side windows. It was filled with nitrogen gas from a gas tank and the temperature inside the chamber was kept at 25 ± 0.5 ºC in the experiment. In addition, in-chamber pressure was well balanced with ambient pressure through a check valve. To avoid vapor condensation hindering the observation, a double layer acrylic window was applied at the front window.

The droplet delivery system, consisting of a syringe pump and a tube guided quartz capillary (300μm ID, 400μm OD), was used to induce nanofluid droplets. A fixed droplet volume of 7 ± 0.1 μL was achieved applying a constant flowrate of 100 μL/min. Though adding nanoparticles slightly alter the surface tension of the fluid, which may result in a difference of droplet size, no pronouncing volume change was observed in the coverage of the present work. The droplet was released at a small Weber number ($We$ ~ 1.0 to give the range) onto the center of the subcooled substrate. Here, $We = \rho u_0^2 R_0 / \gamma$, where $\rho$ is the density of fluids and $u_0$ is the impact velocity.

The high-speed visualization system, consisting a high-speed camera (M310, Phantom) with compact lenses and an optical fiber illuminator, was used to monitor the depositing and freezing process of nanofluid droplets. A grid diffusive glass was applied in front of the illuminator to eliminate reflection and bright points. A fixed capturing speed of 200 frames per second was applied in this experiment.

**Acknowledgements**

We acknowledge the financial support from the Ministry of Singapore via Academic Research Fund (MOE2016-T2-1-114) to CY and the Nanyang Technological University PhD Scholarship to YZ.